\documentclass[10pt,letterpaper]{article}
\usepackage{opex3}

\usepackage{graphicx}
\usepackage{amssymb}

\begin{document}


\title{Complete experimental characterization of a superconducting multiphoton nanodetector}

\author{J.J. Renema\textsuperscript{1)}, G. Frucci\textsuperscript{2)}, Z. Zhou\textsuperscript{2)}, F. Mattioli\textsuperscript{3)}, A. Gaggero\textsuperscript{3)}, R. Leoni\textsuperscript{3)}, M.J.A. de Dood\textsuperscript{1)}, A. Fiore\textsuperscript{2)}, M.P. van Exter\textsuperscript{1)}}

\address{1) Leiden University, Niels Bohrweg 2, 2333 CA Leiden, the Netherlands \\ 2) COBRA Research Institute, Eindhoven University of Technology, P.O. Box 513, 5600 MB Eindhoven, the Netherlands \\ 3) Istituto di Fotonica e Nanotecnologie (IFN), CNR, via Cineto Romano 42, 00156 Roma, Italy }

\email{renema@physics.leidenuniv.nl}

\begin{abstract}
We present a complete method to characterize multiphoton detectors
with a small overall detection efficiency. We do this by separating
the nonlinear action of the multiphoton detection event from linear
losses in the detector. Such a characterization is a necessary step
for quantum information protocols with single and multiphoton detectors
and can provide quantitative information to understand the underlying
physics of a given detector. This characterization is applied to a
superconducting multiphoton nanodetector, consisting of an NbN nanowire
with a bowtie-shaped subwavelength constriction. Depending on the
bias current, this detector has regimes with single and multiphoton
sensitivity. We present the first full experimental characterization
of such a detector.
\end{abstract}




\section{Introduction}

Multiphoton detection is a vital tool for optical quantum computing
\cite{Knill2001}. Such multiphoton detection can take many forms,
two important examples of which are \emph{photon-number resolved detection,
}where the detector is able to distinguish precisely the number of
photons, and \emph{threshold detection}, where the detector is merely
able to distinguish between the cases 'N photons or more' and 'fewer
than N photons' \cite{Feito2009}.

The common factor in all multiphoton detectors is that they are based
on a nonlinear mechanism such that the response of the detector depends
in some nontrivial way on the number of photons impinging on the detector.
There is typically also a finite probability that a photon impinging
on the detector does not participate in the detection process at all.
Such losses can be modeled as attenuation of the input state impinging
on an ideal (i.e. 100\% efficient) detector \cite{Feito2009}.

A well-established tool to characterize any quantum detector is detector
tomography, for which the mathematical framework is that of Positive
Operator Valued Measures (POVM) \cite{Feito2009,Lundeen2008,Akhlaghi2009}.
In this characterization technique, the goal is to find the probability
that the detector clicks, given that N photons are incident on the
detector. These probabilities can be determined by illuminating the
detector with a set of coherent states, and measuring the probability
that the detector clicks as function of the input power. The power
of detector tomography is that it allows us to characterize the detector
using only coherent states as a probe. To do this, it takes into account
the distribution of photon numbers in a coherent state and gives the
probability of the detector responding to N photons.

Without introducting further assumptions, detector tomography is not
immediately applicable in the situation where there is a large and
unknown loss component in the detector. In this regime, the outcome
would be heavily influenced by the probabilities dictated by the linear
losses. To characterize the mulitphoton behaviour of the nonlinear
detection mechanism, the range of test states would have to be large
(of the order $\eta_{sde}^{-1}$, where $\eta_{sde}$ is the system
detection efficiency) which would result in an overwhelming number
of free parameters leading to a strongly overdetermined system.

In this work, we present a method to separate the nonlinear detection
mechanism from the linear loss. We apply this method to the case of
an NbN nanodetector, where we obtain the first full experimental characterization
of such a detector.

This characterization gives the complete statistics of the response of the detector to any incoming state, which is of interest when a detector is used in a quantum communication or quantum information experiment. In addition, since this characterization
is model-independent, it can be used to investigate the physics of the detection mechanism. This latter application is especially important in detectors where the detection mechanism is not fully understood, as is the case for an NbN nanodetector \cite{Hofherr2010}.

\section{NbN Nanodetectors}

NbN Nanodetectors consist of a bow-tie shaped constriction in an NbN
nanowire \cite{Bitauld2010}. The width of this constriction can be
as small as 50 nm. This detector functions on the same detection principle
as the well-known NbN meanders \cite{Goltsman2001}. A detection event
happens when one or more photons induce a break in the superconductivity
and cause the formation of a resistive bridge across the detector
\cite{Hofherr2010}, causing a voltage pulse which is detected by
the readout electronics. In the nanodetector, since the bias current
density is only high around the constriction, this detector has subwavelength
resolution \cite{Bitauld2010}. Furthermore, it is possible to lower
the bias current to such a value that multiple photons are required
to provide a perturbation that is strong enough to break the superconductivity.
Operation in this regime results in a subwavelength multiphoton detector.
Such a detector may allow for subwavelength mapping of optical fields
and high-resolution near-field multiphoton microscopy.

The operation of the NbN nanodetector differs from that of an ideal
N-photon threshold detector, as was already observed in the first
paper announcing the construction of such a device \cite{Bitauld2010}.
In order for these detectors to be used in e.g. subwavelength mapping
of N-photon optical fields, it is vital that their response to different
photon numbers is well understood \cite{Afek2010}.

A complete characterization of the detector may also be of fundamental
interest for the study of the more well-known NbN nanowire meander
detectors \cite{Goltsman2001}. Due to the well-localized sensitive
area of the detector, the multiphoton regime is more apparent and
more easily understood in a nanodetector as compared to an NbN meander,
where two impinging photons are most likely absorbed in different
areas of the detector. Furthermore, it has been suggested that unintended
constrictions form an important limitation on the performance of an
NbN meander \cite{Kerman2007}. In these respects, NbN nanodetectors
may serve as models for the response of NbN meanders.

\begin{figure}[htb!]
\centering
\includegraphics[width=80mm]{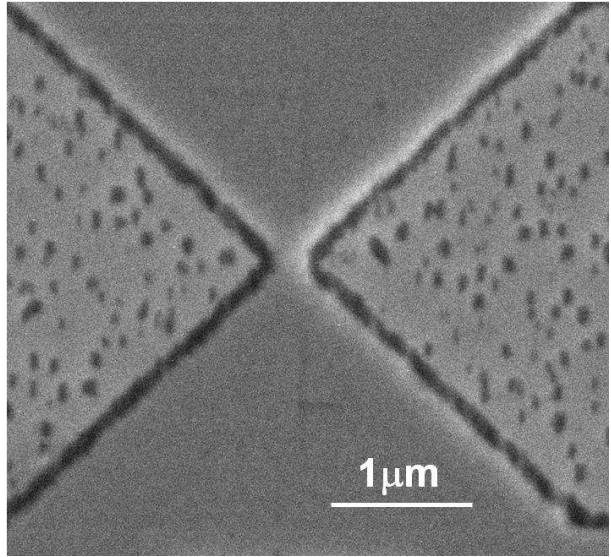}
\caption{SEM image of the NbN nanodetector. The smooth gray area is the NbN,
with the constriction in the middle. From this image, the width of
the constriction was estimated to be 150 nm.}
\end{figure}

\section{Experimental setup}

The NbN nanodetector used in this experiment was manufactured on an
NbN film deposited by DC magnetron sputtering on a GaAs substrate
\cite{Gaggero2010}. A nanodetector was patterned out of the NbN film
by means of Electron Beam Lithography (EBL) and Reactive Ion Etching
technique (RIE). The width of the constriction was estimated to be
150 nm (see figure 1). The detector was cooled in a VeriCold cryocooler
with a final Joule-Thompson stage to 1.17 K. The detector critical
current was measured to be 29 $\mu$A.

We illuminate the sample with a Fianium supercontinuum laser with
a repetition rate of 20 MHz and a pulse width of 6 ps, which was filtered
to have a center wavelength of 1500 nm, with a spectral width of 10
nm. The detector was illuminated through a single-mode lensed fiber
producing a nominal spot size of 3 $\mu$m at 1500 nm. The readout
electronics consist of a bias-tee (Minicircuits ZNBT-60-1W+), an amplifier
chain and a pulse counter.

Each experiment consists of a large series (>20) of experimental runs,
each at constant light power, where the current through the detector
is swept by means of voltage biassing, resulting in steps of 0.2 $\mu$A,
up to the critical current. Power stability during each run was monitored
by a power meter which receives a pick-off beam from a beam splitter
in the fiber leading to the experiment. Finally, the 2-dimensional
set of count rates $C(I_{b},N)$ is rearranged and normalized by the
repetition rate of the laser to yield the detection probability per
pulse $R(N)$ at fixed bias current $I_{b}$.

For each experiment, the power was varied so as to obtain the complete
detector response curve from detection probability $R=10^{-6}$ to
$R=1$. This required varying the input power over 5 orders of magnitude,
typically from 20 pW to 5 $\mu$W input power into the cryostat. At
a repetition rate of 20 MHz the largest input power corresponds to
$N=2*10^{6}$ incident photons per pulse. Since the detection efficiency
of our detector is low (order $10^{-4}$), it was not necessary to
introduce further attenuation, as is usually done in detector tomography
experiments \cite{Feito2009}.

\section{Effective Photon Detector Characterization}

To understand the optical response of the NbN nanodetector, our starting
point is detector tomography, which has been developed in \cite{Lundeen2008,Feito2009,Lundeen}
in the framework of the POVM formalism. This technique provides an
assumption-free method to characterize the response of an unknown
detector system using a set of coherent states as inputs. We limit
ourselves to the case where there are only two possible responses:
click or no click. The idea is to translate the response of the detector
from the basis in which it can be measured (the coherent state basis)
into the basis in which we want to know it, which is the Fock (number)
state basis \cite{Lundeen2008} . For an input state described by
a density matrix $\rho$, the probability $R$ to observe a click
is:
\begin{eqnarray}
R_{click} & = & Tr(\Pi_{click}\rho)\\
\Pi_{click} & = & \sum_{i=0}p_{i}|i><i|,
\end{eqnarray}
where $\Pi_{click}$ is the POVM operator of having a click, and $p_{i}$
is the probability of a click occuring given a Fock state with i photons
as input.

Keeping in mind that for coherent states, the probability distribution
of photon numbers is completely determined by the mean photon number
of the state, we can write:

\begin{equation}
R_{click}(N)=\sum_{i=0}p_{i}c_{i}(N),
\end{equation}
where $c_{i}=e^{-N}\frac{N^{i}}{i!}$ is the weight of the ith basis
state in the probe coherent state and N is the mean photon number.
By measuring the click probability R as a function of the input mean
photon number N of the coherent state, we can use $c_{i}(N)$ to reconstruct
the set of probabilities $p_{i}$, either by a maximum likelyhood
algorithm \cite{Feito2009} or a simple curve fit \cite{Brida2011}.
Since we are dealing with a detector that saturates, i.e. that always
clicks at high input power, the problem is simplified by reasoning
from the case that the detector doesn't click \cite{Amri2010}. Since
there are only two possible outcomes, this gives:
\begin{eqnarray}
R_{click}(N) & = & 1-R_{no\: click}(N)\\
 & = & 1-e^{-N}\sum_{i=0}(1-p_{i})\frac{N^{i}}{i!},
\end{eqnarray}
where N is the mean photon number. The case $p_{0}=0,$ $p_{i>0}=1$
is applicable to any one-photon threshold detector, such as an APD
with unity detection efficiency \cite{Feito2009}.

In this paper, we introduce an extension of detector tomography designed
for use in situations where there is a large linear loss, as is the
case with NbN nanodetectors. The goal of this model, which we call
Effective Photon Detector Characterization (EPDC), is to separate
linear losses from the nonlinear action of the detector, which is
of physical interest. To account for this loss, we introduce a \emph{linear
loss parameter} $\eta$ that describes the probability of for each
photon to participate in the nonlinear process. Since coherent states
remain coherent under attenuation, the EPDC function then becomes:

\begin{equation}
R_{click,EPDC}(N)=1-e^{-\eta N}\sum_{i=0}(1-p_{i})\frac{(\eta N)^{i}}{i!},
\end{equation}
where $\{p_{i}\}$ and $\eta$ are the free parameters. Since the
POVM description is complete \cite{Lundeen2008,Amri2010} and we have
added a parameter, we have now created a function that is overdetermined
by one parameter. However, we can choose a solution based on physical
grounds. Since we know our detector has threshold-like behaviour,
it is reasonable to assume that for some large number of photons $i_{max}$
the probability $p_{i_{max}}$ with which the detector will click
is arbitrarily close to 1. Furthermore, once we have found such an
$i_{max}$, we can assume that $p_{j>i_{max}}=1$ for all $j>i_{max}$,
since otherwise we would have the unphysical case that adding photons
makes it less likely that the detector clicks. We can then create
a series of candidate solutions by fitting Eq. 6 to our measured count
rates as a function of input photon number, truncating the sum at
various values $i_{max}$. This gives a series of candidate solutions
parameterized by $\{\eta,\: p_{0}...p_{i_{max}}\}$. The solution
we pick is the one that fits our data and has the minimum $i_{max}$,
since this is the one that requires the fewest parameters to explain
our data.

The big advantage of this approach is that we describe all the linear
loss with a single parameter, thereby separating the linear losses
from the nonlinear action of the detector, and drastically reducing
the number of fit parameters. Typically, the nonlinear action of the
detector, quantified by the $p_{i}$ is the quantity of interest for
multiphoton detection. This approach is particularly relevant for
detectors with a large linear loss component, since if this loss is
not taken into account separately it would dominate the characterization
of the detector.

\section{Result}

The points in Fig. 2 show the measued count rate points as a function
of input power from our NbN nanodetector at three different bias currents.
The lines represent the fits, with the colour indicating the value
of $i_{max}$ (see legend). For each fit the reduced $\chi^{2}$ are
shown in the bar diagrams in the insets of the figure. We take the
fit that explains the data with the smallest number of parameters
as the most physically realistic solution. This choice is indicated
by the arrows in the bar diagrams. By repeating this algorithm over
a range of bias currents, we can completely characterize how the response
of the detector to a given number of photons varies with the bias
current.

In figure 3, the results from the Effective Photon Detector Characterisation
are shown as a function of bias current. At each bias current, the
obtained $p_{i}$ and $\eta$ describe the operation of the detector,
independent of power. We therefore conclude that we have obtained
a complete description of the detector behaviour.

\begin{figure*}[htb!]
\centering
\includegraphics[width=120mm]{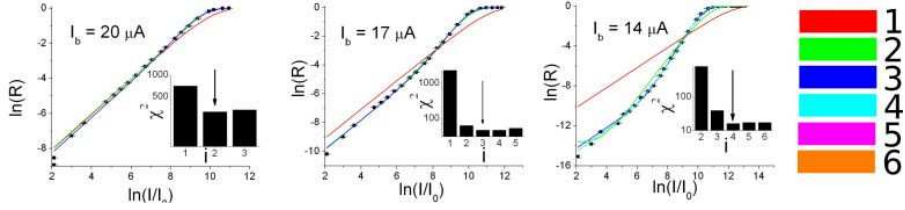}
\caption{Measurement of the NbN nanodetector count rate as function of input
power at Ib = 20 $\mu$A, Ib = 17 $\mu$A and Ib = 14 $\mu$A, fitted
with the EPDC model (Eq. 1). The black squares represents the data
points with error bars, the other lines represent fits, with the number
of free parameters represented by the color of the line (see legend).
Note that many of these lines overlap with each other and with the
data. Insert: reduced $\chi^{2}$ of the fits as a function of number
of parameters. For $I_{b}=14\mu A$ we have omitted the case $i=1$,
where $\chi_{reduced}^{2}>10^{4}$. The arrows indicate the best fit.
Note that in all three cases there are multiple fits which have similar
reduced $\chi^{2}$, we reject the ones with superfluous free parameters
for physical reasons. }
\end{figure*}

\begin{figure}[htb!]
\centering
\includegraphics[width=120mm]{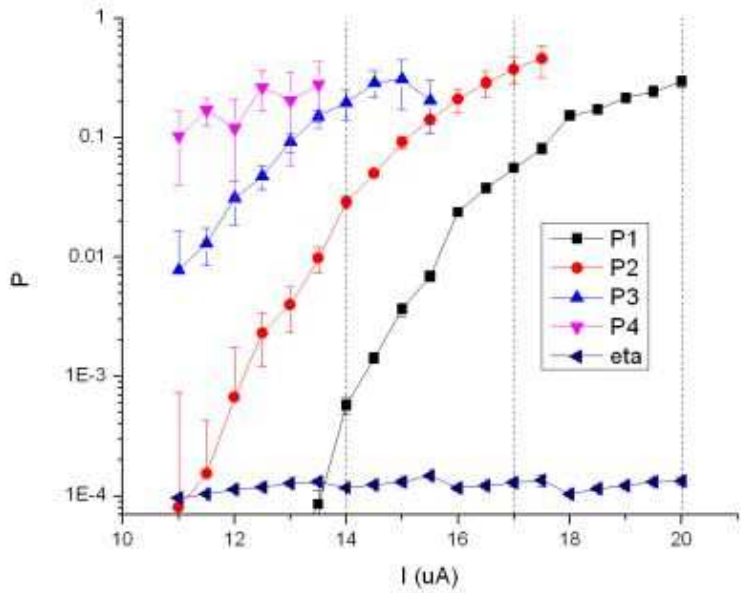}
\caption{EPDC parameters and linear detection efficiency as function of bias
current. This figure was obtained by repeatedly applying the method
outlined in Section IV at various bias currents. The three dashed
lines indicate the bias currents from Figure 3. }
\end{figure}

\section{Discussion}

The $p_{i}$ obtained from the fit represent the nonlinear action
of our detection system, which is the physical property of interest.
Since there are no other nonlinear elements in the detection system,
we can unambiguously attribute the behaviour of the $p_{i}$ to the
NbN nanodetector. It should be noted that result presented here is
consistent with earlier results on these detectors \cite{Bitauld2010},
e.g. we reproduce the finding that the transitions between the various
detection regimes (where the detector behaves approximately as an
N-photon detector) are equally spaced in the current domain.

From Eq. 3, we can see that the response of the detector is given
by terms of the form $p_{i}c_{i}(N)$, where $c_{i}(N)$ is the probability
of having N photons. From this we can see that each $p_{i}$will be
most dominant in the range of powers where the probability of having
the corresponding number of photons is highest. For example, at 17
$\mu$A the detector has $p_{1}$= 0.06 and $p_{2}$= 0.37, meaning
that at low powers ($\eta N<0.16$), where the one-photon contribution
from the state is dominant, the detector will respond mostly to single
photons, but at higher powers ($\eta N>0.16$) the response will be
dominated by the two-photon events. This quantifies the change of
detection regimes reported in measurements of count rate as a function
of power \cite{Bitauld2010}.

The fitted linear detection efficiency $\eta$ fluctuates between
$(9.6\pm0.2)*10^{-5}$ and $(14.7\pm0.6)*10^{-5}$. Normalizing to
the estimated effective area of the detector of 100 nm by 150 nm and
the beam size, we obtain an intrinsic detection efficiency of 8\%.
While it should be noted that this is only a rough estimate, it is
higher than the value of 1\% reported in \cite{Bitauld2010}, we attribute
to the lower temperature of the experiment, at which NbN detectors
are known to be more efficient \cite{Hofherr2010}.

It should be noted that since we combine all linear losses into a
single parameter, we are unable to distinguish losses after the absorption
event from those before the absoption event, provided they are linear.
It is known for NbN meanders that not every absorbed photon causes
a detection event \cite{Kerman2007}. However, since our measured
linear loss does not depend on the bias current, it is reasonable
to attribute it to optical loss and not to losses inside the detector.
With the caveat that there may be a constant linear loss inside the
detector, we can therefore conclude that the set of $p_{i}$ completely
describes the behaviour of the detector.

\section{Conclusion}

We have introduced an extension of detector tomography which is applicable
in the presence of a large linear loss. This detector characterization
is of interest when using a quantum detector in a quantum optics or
quantum communication experiment, since it gives a full prediction
of the response of the detector to any incoming state. Furthermore,
we have completely characterized the response of a superconducting
nanodetector, over several operating regimes of the detector. This
represents the first complete characterization of this type of detector, which is necessary for the use of this detector as a  multiphoton subwavelength detector.

A second application of this characterization method is that it provides quantitative information about the response of the detector. Such quantitative characterization can also be used to test theoretical predictions of the response of the detector as a function of bias current \cite{Semenov2005}, enabling further insight into the physics of the detection event in NbN photodetectors.

The idea of our formalism is to separate this linear loss from the nonlinear action of the detector. For the detector under study in
this paper, this formalism completely describes the response of the detector. In contrast to earlier methods \cite{Akhlaghi2009a,Divochiy2008} that assume a priori knowledge of the underlying detector physics,
detector characterization based on the POVM formalism can be applied to any detector system without making assumptions about the operating principle of the detector \cite{Lundeen2008,Akhlaghi2009}. Therefore, the strength of the characterization applied here is that we can extract model-independent parameters that can be used to gain insight in the physics of photon detection with NbN detectors.

Finally, we comment on the applicability of our algorithm to other
detectors: Effective Photon Detector Characterization shares the feature
with detector tomography that it is as assumption-free as possible;
making it possible to characterize a detector without any prior knowledge
or model of the operational mechanism of the detector. The EPDC method
has the added requirement that the detector saturates (i.e. always
produces the same outcome) at some high input photon number. To our
knowledge, this behaviour is generic to all quantum detectors constructed
to date \cite{Haderka2003,Rohde2007,Divochiy2008,Dauler2008}.
It therefore does not represent an practical limitation.

\section{Note}

This work is part of the research programme of the Foundation for
Fundamental Research on Matter (FOM), which is financially supported
by the Netherlands Organisation for Scientific Research (NWO). JR
would like to thank S. Jahanmiri Nejad for experimental advice.

\end{document}